\documentclass[11pt]{article}
\usepackage{graphicx}
\title{\bf On the Way to Submicroscopic Description of Nature\footnote
{{\it Indian Journal of Theoretical Physics} {\bf 49}, no. 2, pp.
81-95 (2001);  \ \ \ \ \ \ \quad \quad \quad \quad \quad  \break
also arXiv.org e-print archive \
http://arXiv.org/abs/quant-ph/9908042}}

\author{{\bf Volodymyr Krasnoholovets}  \\
 {} \\
Institute of Physics, National Academy of Sciences, \\ Prospect
Nauky 46,   UA-03028 Ky\"{\i}v, Ukraine \\ http://inerton.cjb.net}

\date{4 August 1999}

\begin{document}
\maketitle

\begin{abstract}
The outline analyzes the principal difficulties, which emerge at
the applying of modern quantum theory based on the Copenhagen
School concept to phenomena developed in the range close to
$10^{-28}$ cm (the point of intersection of the three fundamental
interactions). It is shown that at this scale, the interaction of
a moving particle with space plays an essential role and just
space assigns wave and quantum properties to the particle. The
main physical aspects of space structure are discussed herein.

\vspace{2mm}


\end{abstract}

\vspace{5mm}

\section{Question-marks germinating out of basement}

\hspace*{\parindent}
 Modern physics axiomatic is constructed on a very abstract mathematical
formalism that is aimed only at the quantitative description of physical
phenomena. For instance, the Maxwell equations describe electromagnetic
phenomena but they do not bring an idea of the structure of the charge
and the electromagnetic field. The Schr\"odinger equation describes
quantum mechanics of particles but the equation cannot explain the
reason of long-range action and wave behavior of the particles.
On the question what is photon?, quantum electrodynamics answers:
it is something that can be described by the equation
\[
\partial^{{\kern 1pt}2} {\vec A} / \partial {\kern 1pt}
t^{{\kern 0.5pt}2} - c^{-2} {\kern 1pt}
\partial^{{\kern 1pt}2} {\vec A} / \partial {\kern 1pt}
{\vec r}^{{\kern 2pt} 2} =0
\]
where $\vec A$ is a physical value called the "vector potential".

    Such a structure of the formalism does not permit to reveal the
origin of processes constituted the essence of quantum phenomena
studied. Lorentz also pointed the same in the beginning of 20th
century (see, e.g. Ref. [1]). De Broglie held the viewpoint [2]
that there are hidden laws (see also Bohm [3]), which provides the
basis for motion and that the description of phenomena should also
be the goal of physics, not only their prediction. Nonetheless, it
is now believed that so-called "unorthodox" questions are
irrelevant. However, can we correctly understand the behavior of
elementary particles, if the whole series of fundamental notions
researchers operate with everyday have not become clear yet? For
example, one can raise the following questions.

\vspace{1.5mm} \noindent 1)  What is $\psi$-wave function? This
problem still thrills the curiosity of researchers (see, e.g.
review [4]).

\vspace{1.5mm} \noindent 2)  All correct theories should be
Lorentz invariant, i.e. they and Einstein's special relativity
should agree (see, e.g. Ref. 5). Nevertheless, the Schr\"odinger
equation is not Lorentz invariant but it perfectly describes
quantum phenomena. How is it possible?

\vspace{1.5mm} \noindent 3)  Why does the classical parameter $M$
-- the particle mass -- enter the Schr\"odinger quantum equation?
Where is the particle mass when the particle as the whole is
fuzziness in an undetermined volume as the  $\psi$-wave function
prescribes?

\vspace{1.5mm} \noindent 4)  What is mass? In modern quantum field
theories mass is considered as a characteristic expressed through
the energy $E$ and momentum $p$. Today theorists try to assure
[6-8] that mass does not depend on velocity. And this is very
strange because it turns out from such determination that the
notion of mass can not be considered as a quantity of matter which
is found in the volume of a particle/body. Moreover such
declarations are in contradict to the experiments by Bucherer [9]
and Rogers {\it et al.} [10] who studied the dependence of mass on
the velocity and confirmed the validity of the formula
$M=M_0(1-v^2/c^2)^{-1/2}$.

\vspace{1.5mm} \noindent 5)  What are microscopic processes
changing the geometry of space surrounding an object, which
manifest themselves in the form of the Newton/Coulomb potential
$1/r$? Einstein noted [11] that the geometry employing in general
relativity (the Riemannian geometry) should be treated only as a
macroscopic geometry. In other words, what is origin of the
gravity?

\vspace{1.5mm} \noindent 6) There is no correct determination of
values $E$ and $\nu$ in the expression $E = h \nu$ applied to a
moving canonical particle.  In one case $E=\frac 12 M_0{\kern
1pt}v^2$ (see, e.g. Ref. [12]), and in the other one
$E=M_0c^2(1-v^2/c^2)^{-1/2}$ (see, e.g. Ref. [13]).  Which is
true?

\vspace{1.5mm} \noindent 7)  The description of a quantum system
in terms of the Dirac field or Dirac's equation is correct only at
the scale $r\geq \lambda_{\rm Com}$ where $\lambda_{\rm Com}$ is
the Compton wavelength. Hence, what a physical characteristic of
space in the vicinity of particle does the parameter $\lambda_{\rm
Com}$ describe? And what approach can make used at the scale $r<
\lambda_{\rm Com}$?

\vspace{1.5mm} \noindent 8)  What is spin? It is one more mystery
of the microworld. Quantum field theories define it as an
"inseparable and invariable property of a particle" [14]. That is
all.

\vspace{1.5mm} \noindent 9)  What is nature of the phase
transition that turns us from the description of a quantum system
based on the Schr\"odinger equation to that based on the Dirac
one?

\vspace{1.5 mm} \noindent 10) What is nature of the fundamental
physical  constants $c$, $h$, and $G$? If the value of $c$ is
constant then why does the experiment register the superluminal
velocity (from $c$ to $4.7 c$) [15]?

\vspace{1.5mm} \noindent 11)  What is electric charge and why is
it fractional in quarks?  It is said that the charge is something
that is written in quotation marks [16]; constants of gauge
interactions are called the "charge" as well.  Thereby, according
to the definition [17] the electric charge is a value that is
measured by the elementary electric charge unit $e$!

\vspace{1.5 mm} \noindent 12) What is structure of real space?

\vspace{2mm} It is easily seen that answers to these questions
cannot be found solely in the framework of power mathematical
methods of contemporary physics. Notwithstanding this, the
questions are pure physical and we should look for the solvability
of all these very urgent problems of fundamental science. We
should answer what the canonical particle is? What is its size?
And what all the properties mentioned above do mean?

\vspace{4mm}

\section{Confusions of quantum theory}

\hspace*{\parindent}
     Present views on the canonical particle are restricted by the
following primitive notion that is not reconcilable with a very
difficult and formal mathematical construction, which is applied
to the description of the quantum systems behavior. First,
experimental data correspond to the length $l\geq 10^{-17}$ cm
[18]. At this scale we should imagine a "black box" that unified
our perception about the particle. The box is pasted over various
labels, which contain legends like these: the mass is equal to
$M$, the spin is equal to 1/2, the energy $E$ is equal to $h\nu$,
the charge is equal to $e$, and so on. In the case of quarks,
things get worse: an isolated quark does not exist and that is why
we can talk about the box only resorting to indirect information
on the spin, color and so on, that is the notions which need
submicroscopic investigations themselves.  Second, we should
rather substitute the notion of fundamental particle (i.e. our
abstract black box) by the more abstract notion of fundamental
symmetry [19]. Third, at the atom size the fundamental symmetry is
suddenly transformed into the $\psi$-wave function or spinor
$\hat\psi$, as the case requires.  The case is a function of the
ratio $v/c$ where $v$ and $c$ are the particle and light velocity
respectively.  It turns out that the $\psi$-wave function and
spinor $\hat \psi$ can be not only considered as the fundamental
symmetry parameters but parameters of the particle as well.  At
the same time special relativity says that the value $v_0$ is not
absolute and depends on a frame of reference.  One may choose such
a frame of reference that $v_0$ will be very close to $c$. This
means that the Schr\"odinger formalism may be easily replaced for
the Dirac one and on the contrary, we can choose such frame of
reference for a quantum system described by the Dirac equation
that the Dirac formalism will smoothly pass to Schr\"odinger's.
However everybody knows that this is absurdity and moreover at
this point an internal inconsistency of the theory comes to light:
in the Schr\"odinger quantum equation the distance between two
instantly interacting electrons, as was noted by Ehrenfest [20],
"can be equal to any quantity of kilometers". Besides the
Schr\"odinger equation is not Lorentz invariant and therefore
formalism based on this equation can not be conjugated with that
resting on the Dirac one, whereas both the formalisms are
confirmed by experiment perfectly.  So such a strange theory we
have.

    On the other hand, there are exact postulates, which directly
follow experimental facts. First, there are corpuscles whose
behavior similar to wave. Second, the velocity and mass are
characteristics of objects; the size of objects contracts in the
direction of motion and objects' masses increase with velocities.
Third, two basic quantum mechanical relations are applied for any
particle: $E=h\nu$ and $\lambda=h/Mv$. Forth, each particle has
its own limiting length, the Compton wavelength $\lambda_{\rm
Com}$, behind of which quantum fluctuations of the vacuum are
absent.  Fifth, there is a quantum characteristic of the particle
called the spin that can contribute the orbital momentum of the
particle. Sixth, when the velocity  $v$ of the particle approaches
to  $c$, the phase transition takes place in the quantum system
studied and one should pass from the Schr\"odinger formalism to
the Dirac one.

    Loud disagreements in quantum theory point towards the need for
its improvement. All modification must keep pace with reliable
established experimental facts. To solve the problem we should try to
study the three following subjects together, which have never been
previously considered as the whole: Foundations of quantum mechanics,
Foundations of quantum gravity, and Foundations of quantum electricity.
But the first point of the study is the structure of the geometry of
space and a correct definition of real space.

\vspace{4mm}

\section{Search for a submicroscopic approach}

\hspace*{\parindent} There are different approaches to the
problem. Among them one can name a new approach by Hofer [21] who
has proposed to consider electrons and photons as extended
particles which comply with a wave equation; the newest concept
offered by Kirilyuk [22], who has constructed a theory of two
fundamental fields, which lead to the universal concept of dynamic
complexity and the permanently developing hierarchical structure
of the universe. Kiriliuk's model demonstrates a possibility of
the double solution with chaos, which takes into account the
deterministic concept of quantum mechanics developed by de
Broglie. Based on causal interpretation of quantum mechanics
pioneered by de Broglie [23] and Bohm [24], Roy and Singh [25]
have suggested a deterministic mechanics in which the quantum
probability densities are simultaneously reproduced as marginal of
one positive defined phase space density, which is constant along
the trajectory. We shall point out also several other new views on
the nature of a vacuum and real space which have appeared [26-32]
during the last decade. These see a vacuum as a substance and
determine matter as deformations of space, Bounias and Bounaly
[31,32]; in particular, papers [31,32] studied premises for the
existence of an initial cell of space in terms of the topology and
the set theory. About some kind of a primordial cell and existence
principle was also pointed in Ref. [33].

     In high energy physics wave properties of canonical particles are
neglected and the behavior of a quantum system is often described
drawing an analogy with the lattice model, string model, bubble
model, bag model, etc. Thus, in many cases quantum field theory
can not preclude ideas and concepts used in condensed media
physics. Because of this account must be taken of the
microstructure of the vacuum, i.e., real space, at the scale of
the order of  $10^{-28}$ cm (at this size electromagnetic, weak
and strong interactions come together). It is likely that space at
this scale can be simulated as an order/disorder lattice, similar
to a solid/liquid, or as a cellular structure, similar to the pack
of soft spheres. (It is interesting to note that such view is
conceptually close to the re-introduction of some kind of an
aether, but the quantum one.)

     In the author's works [34-36] real space has been simulated in the
form of a quantum substance as well and an elementary cell of
space -- a superparticle -- has been offered. The origin of
matter, a local space curvature, or deformation, is created when
the volume of an initial cell changes. One can consider the local
deformation as a corpuscle. Let us look now if the model can
explain quantum mechanics phenomena. How can the local deformation
that can be treated as a corpuscle whose size is limited by
$10^{-28}$ cm moves similar to a wave and manifests the wave
behavior at the atom scale? Of course, it is not an easy problem
but it has a solution. The motion of the physical "point"
(corpuscle cell) in entirely packed discrete space must be
accompanied by the interaction with the "points" of space
(superparticle cells) giving rise to excitations in neighboring
cells. Note that the similar phenomenon occurs in a solid: a
particle moving in the solid brings about excitations such as
excitons, solitons, etc. As the excitations are associated with
the motion of the corpuscle they were called "inertons", i.e. just
the corpuscle inert mass is responsible for the creation of such
kind of space excitations. A portrait of the moving corpuscle is
depicted in Fig. 1.
\begin{figure}
\begin{center}
\includegraphics[scale=2]{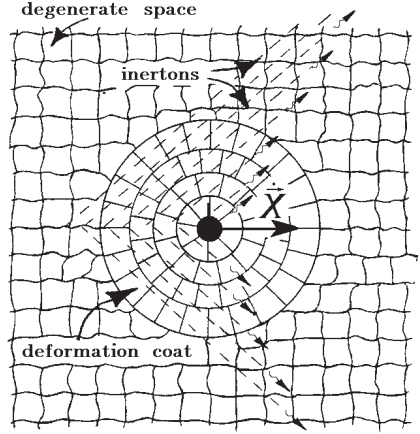}
\caption{The corpuscle, i.e. canonical particle (\textbullet),
moving in the space net.} \label{1}
\end{center}
\end{figure}
When the corpuscle moves it also pulls its deformation coat, in
other words, the space crystallite. The crystallite migrates by a
relay mechanism: in any place of corpuscle location surrounding
cells are made ready as the Figure demonstrates. The crystallite
is similar to a shell that screens the corpuscle from degenerate
space. Cells have mass inside the crystallite and have not it out.

   The dynamics of the corpuscle and its inerton cloud was studied
in Refs. [34-36]. It was shown that the cloud of inertons
oscillates around the corpuscle with amplitude $\Lambda$ that can
be found using the relationship
\begin{equation}
\Lambda=\lambda{\kern 1pt} c/v_0 \label{1}
\end{equation}
where $\lambda$ is the amplitude of spatial oscillations of the corpuscle,
$c$ is the speed of light and $v_0$ is the corpuscle's initial velocity.

    It is quite reasonable to center the notion of corpuscle,
i.e., local deformation of space, on the notion of canonical
particle. This enables [34,35] to make a deeper microscopic
interpretation of the de Broglie wavelength $\lambda$ as amplitude
of spatial oscillations of the particle along its path: on the
first half of spatial period of oscillations $\lambda/2$ the
particle emits inertons and its velocity decreases from $v_0$ to
0. The emitted inertons gradually retarded by elastic cellular
space. Then space returns the inertons to the particle and hence
on the second half of the period $\lambda/2$ the particle absorbs
the inertons and its velocity increases from 0 to $v_0$, and so
on. It is significant that the inertons emitted ahead of the
particle are absorbed behind of it (owing to the difference
between the inerton and the particle velocities, $c>v_0$). The
submicroscopic mechanics of the particle permits to deduce
correctly the basic quantum mechanical relations:
\begin{equation}
E=h\nu,  \ \ \ \ \ \ \ \ \    \lambda= h/p;
\label{2}
\end{equation}
here, in our case $p=Mv_0$ and $M=M_0(1-v_0^2/c^2)^{-1/2}$.
Relations (2) were not obtained previously from any theory; they
were only postulated by de Broglie who showed [37] that the
Schr\"odinger equation is their consequence. However, unlike the
traditional presentation, the Schr\"odinger equation gained in
paper [35] is Lorentz invariant because it includes time $t$ as a
natural parameter that is Lorentz invariant by definition.

   Space covered by the inerton cloud determines the range of the
wave $\psi$-function action. The cloud of inertons accompanying
the moving particle contacts any obstacles around the particle in
a distance $\sim \Lambda$ and transmits a respective information
to the particle and this is the easiest explanation of the
particle diffraction phenomenon. The size of the deformation coat,
or the crystallite is equal to the Compton wavelength
$\lambda_{\rm Com}$; it is just the parameter that characterizes
the relativistic behavior of the particle, in particular, the
photon scattering by it.

    What is inerton? It is a quasi-particle that carries an elementary
deformation from cell to cell by a relay mechanism like the
Frenkel (molecular) exciton transferring the energy in a molecular
crystal. Owing to the comparison of the deformation with mass (see
Refs. [34,35]) it is reasonable to assume that inertons should
substitute for hypothetical gravitons -- carriers of the
gravitational interaction of general  relativity. Indeed,  i)
inertons have mass (gravitons have not);  ii) inertons are a part
of any quantum and classical physical system (gravitons were
deduced only from the pure classical behavior of objects and these
particles can not be introduced in quantum mechanics in
principle); iii) inertons can be easily revealed in any physical
laboratory by means of many different tools (the existence of
gravitons has never been confirmed).

    The general theory of relativity did not take into account the
existence of matter waves, which quantize space at the microscopic
scale. Therefore the macroscopic requantization of space that
general relativity predicts is highly conjectural. Moreover the
relativity lumped together real space with time which is a
non-geometric parameter. This is the amalgamation that builds up
enormous obstacles on the way to a microscopic consideration of
the gravitation phenomena. The problems of the construction of a
mathematical space and time emergence in it has recently been
raised by Bonaly and Bounias [31,32]; they have shown that time is
associated with the mapping of intersections of topological
spaces. A detailed theory of space conforming to the experimental
results in the fields of microscopic and macroscopic phenomena is
stated in Ref. [38].

     The impact of inertons on the structure of test specimens has been
demonstrated in paper [39] (other manifestations of inertons are
described in Refs. [40,41]). At the same time many other
physicists have observed unusual effects, which may be caused by
inertons as well. In particular, {\it Europhys. News} has reported
about one of them [42]: a large group of researchers could observe
the electron wave $\psi$-function on metal surface. Nonetheless,
everybody knows that the wave $\psi$-function is just the
mathematical function like to the Boltzmann function $f(\vec r,
\vec{\dot r})$, the Hamilton-Jacobi function $S(\vec r, E)$, etc.
All these do not act in real space and only set connections
between particle's parameters. This is why spherical and
elliptical images showed in the figures in Ref. [42] should be
interpreted as images of inerton clouds surrounding electrons.

     The concept of cellular degenerate space and the submicroscopic
mechanics, which are progressing allow us to disclose many
significant details of the microworld and it is valid to say that
the first significant result obtained in the framework of the
concept is the solution of two difficult problems of
nonrelativistic quantum mechanics. First, the theory developed in
Refs. [34,35] removed a very unpleasant conflict that took place
between nonrelativistic quantum mechanics and special relativity:
Unlike the traditional presentation, the Schr\"odinger equation
gained in paper [35] is Lorentz invariant owing to the invariant
time entered in the equation. Second, due to inertons introduced
in the quantum system nonrelativistic quantum mechanics no longer
suffers from long-range action.

    Such submicroscopic approach is able to give us great insight
into both the submicroscopic structure of canonical particles and
the particles dynamics, which are hidden from observation inside
the "black box" that presents an impenetrable barrier to the
quantum field theory, string theory, supersymmetry, supergravity,
and others. The theory based on the concept of fine-grained
degenerate space is a newest one, however, it would be the
shortest way to the unified theory of matter. The new concept
needs new mathematical ideas, new approaches and a new research
methodology.

\vspace{4mm}

{\bf Acknowledgement}

\vspace{2mm}

\hspace*{\parindent} I am very thankful to Professor Michel
Bounias for fruitful discussions and valuable remarks.

\end{document}